\documentclass[aps,prl,twocolumn,showpacs]{revtex4-1}
\usepackage{graphicx}
\usepackage{bm}
\usepackage{amsmath}
\usepackage{amssymb}

\providecommand{\ket}[1]{\lvert #1 \rangle}

\begin{document}

\title{Fractional topological phase for entangled qudits}
 
\author{L.~E.~Oxman and A.~Z.~Khoury}

\affiliation{
Instituto de F\'\i sica, Universidade Federal Fluminense,
24210-346 Niter\'oi - RJ, Brasil.}

 
\begin{abstract}
We investigate the topological structure of entangled qudits 
under unitary local operations. 
Different sectors are identified in the evolution, 
and their geometrical and topological aspects are analyzed. 
The geometric phase is explicitly calculated in terms of the 
concurrence. 
As a main result, we predict a fractional topological phase for 
cyclic evolutions in the multiply connected space of maximally 
entangled states. 
\end{abstract}
\vskip2pc 
 
\maketitle

In a seminal work, M. Berry \cite{berry} showed the important 
role played by geometric phases in quantum theory. Since then, 
the interest for geometric phases was renewed by potential 
applications to quantum computation. The experimental demonstration of 
a conditional phase gate was provided both in Nuclear Magnetic 
Resonance (NMR) \cite{vedral} and trapped ions \cite{zoller}.
Optical geometric phases have already been discussed 
both for polarization \cite{pancha} and vortex mode
transformations \cite{vanenk,galvez}. 
The role of entanglement in the phase evolution of qubits was investigated 
in refs.\cite{sjoqvist,sjoqvist2}. 
Recently, P. Milman and R. Mosseri \cite{remy,milman} investigated the 
geometric phase and the topological structure associated with cyclic 
evolutions of arbitrary two-qubit {\it pure} states.  
This structure has been experimentally 
evidenced in the context of spin-orbit mode transformations of 
a laser beam \cite{nossoprl} and in NMR \cite{nmr}. 
Although the topological nature of the phase acquired by maximally entangled states 
is well settled, the distinction between geometrical and topological phases has not 
been established clearly for partially entangled states.  
In this work we present a group theoretical approach which allows for a 
clear distinction between the two aspects. As a bonus, this approach is 
easily extended to higher dimensions, bringing an interesting prediction of a 
fractional topological phase. 

Let $\ket{\psi}=\sum_{i,j=1}^{d} \alpha_{ij}\ket{ij}$ be the most 
general two-qudit pure state. We shall represent this state by the 
$d\times d$ matrix $\mathbf{\alpha}$ whose elements are the coefficients 
$\alpha_{ij}$. With this notation the norm of the state vector becomes 
$\langle\psi|\psi\rangle=Tr(\mathbf{\alpha^{\dagger}\alpha})=1$ and 
the scalar product between two states is 
$\langle\phi|\psi\rangle=Tr(\mathbf{\beta^{\dagger}\alpha})$, where 
$\mathbf{\beta}$ is the $d\times d$ matrix containing the coefficients 
of state $\ket{\phi}$ in the chosen basis. 
We are interested in the phase evolution of the state $\ket{\psi}$ 
under local unitary operations. So let us take two unitary matrices $U_A$ and 
$U_B$ belonging to $U(d)$ and representing the operations performed in each 
subsystem separately. 
Under these unitary operations the state matrix will evolve as 
$\mathbf{\alpha}(t)=U_A\,\mathbf{\alpha}(0)\,U_B^{\intercal}\,$, 
where $U_j(t)=e^{i\phi_j(t)}\bar{U}_j(t)$ ($j=A,B$) and $\bar{U}_j \in SU(d)$.
One can identify the following invariants under local unitary evolutions: 
$Tr[\rho_j^{\, p}]$, $p=1,\dots, d$, where $\rho_j$ is the reduced density matrix 
with respect to qudit $j$ 
($\rho_A =\alpha^{\intercal}\alpha^\ast$ and $\rho_B=\alpha \alpha^\dagger$). 
In fact, the invariants are $j$-independent. The first one ($p=1$) is simply the 
norm of the state vector. One can readily relate the second invariant to the 
{\it I-concurrence} of a two-qudit pure quantum state \cite{Iconc} 
${\cal C}=\sqrt{2(1-Tr[\rho^2])}$, so that its invariance expresses the well 
known fact that entanglement is not affected by local unitary operations. 
The $p=d$ invariant can be rewritten in terms of the former and 
${\cal D}=|\det [\alpha]|$. In particular, for qubits we have 
${\cal C}=2\, {\cal D}$. 

In the case of a cyclic evolution, 
$\bar{U}_A(\tau)\mathbf{\alpha}(0)\bar{U}^{\intercal}_B(\tau) = 
e^{i\Delta\phi}\mathbf{\alpha}(0)\,$. 
By taking the determinant of both sides we get: $e^{i\,d\Delta\phi}=1$ 
as long as ${\cal D}\neq 0$. This implies that the possible acquired phases 
due to the $SU(d)$ part of a cyclic evolution are $\Delta\phi=2\pi n/d$, with 
$n = 0, 1, 2, ..., d-1$. For qubits ($d=2$) one recovers the well known 
result $\Delta\phi=0,\pi$. However, for $d>2$ one 
obtains fractional phase values in steps of $2\pi/d\,$. 
Now, we are interested in discussing in what sense this fractional phase can be 
considered as topological. For this aim, we will analyze the topology of the 
space of two-qudit states and how the total phase is built. In this regard, 
we would like to underline that according to ref. \cite{mukunda}, the geometric 
phase acquired by a time evolving quantum state $\mathbf{\alpha}(t)$ is always 
defined as
\begin{equation}
\phi_g = \arg{\langle\psi(0)|\psi(t)\rangle} + 
i\int dt \,\,\langle\psi(t)|\dot{\psi}(t)\rangle \;,
\label{phig}
\end{equation}
that corresponds to the total phase minus the dynamical phase.
Therefore, a topological phase, that is, an object that only depends on a given 
class of paths, can only find room as a part of the geometric phase, an object 
that is invariant under reparametrizations and gauge transformations. Gauge 
invariance corresponds to the fact that the phase factors $\phi_j(t)$ do not 
contribute to $\phi_g$, which is completely determined by $\bar{U}_j(t)$, 
the sector where the fractional values occur. 

In order to characterize the space of states, we note that any invertible matrix 
admits a polar decomposition
$\alpha=Q\,S$, where $Q=\sqrt[d]{{\cal D}}\,e^M$ is a positive definite Hermitian matrix, 
$M$ is a traceless Hermitian matrix, and $S= e^{i\phi}\,\bar{S}$, $\bar{S} \in SU(d)$. 
Since $\det[e^M]=e^{Tr[M]}=1$, one easily finds 
$\det(\alpha)={\cal D}\,e^{i\,d\,\phi}$. We can identify the time evolution 
as occurring in different sectors
\begin{equation}
\alpha (t)=\sqrt[d]{{\cal D}}\,e^{i\phi(t)}\,e^{M(t)}\,\bar{S}(t)\;,
\label{polarevolution}
\end{equation}
where we have denoted, $\phi(t)=\phi(0)+\phi_A(t)+\phi_B(t)$, 
$M(t)=\bar{U}_A(t)M(0)\bar{U}_A(t)^{\dagger}$, 
and $\bar{S}(t)=\bar{U}_A(t)\bar{S}(0)\bar{U}^{\intercal}_B(t)$.
Therefore, we identify the evolution in three  
sectors of the matrix structure: an explicit phase evolution 
$\phi(t)$, an evolution closed in the space of traceless Hermitian 
matrices $M(t)$, and the evolution $\bar{S}(t)$ closed in $SU(d)$. 

Now we are able to discuss the topological aspects of the entangled 
state evolution in terms of these sectors. The space of 
positive definite Hermitian matrices $Q$ has trivial topology. This is a 
noncompact manifold isomorphous to $R^{d^2-1}$, as it can be parametrized 
in the form $Q=e^{\beta_{a}\,T_{a}}$, where $\beta_{a}$ 
are real numbers and $T_{a}$ ($a=1,2,...,d^2-1$) is a basis in the space of 
Hermitian traceless matrices. 
These $T_a$'s are the generators of $SU(d)$, so that 
$\bar{S}=e^{i\,\omega_{a}\,T_{a}}$, with $\omega_{a}$ real. They can be 
normalized in the form $tr\, (T^a T^b)=\frac{1}{2}\delta^{ab}$ and obey 
the Lie algebra $\left[ T^{a},T^{b}\right]=if^{abc}T^c$, where $f^{abc}$ 
are the structure constants of $SU(d)$.
The first homotopy group of $SU(d)$ is also trivial, however, the physical 
equivalence of $\alpha$ matrices differing by a global phase corresponds to 
considering the identification in $SU(d)$, 
$e^{i 2\pi n/d}\,\bar{S}\equiv \bar{S}$. 
This can be naturally implemented by associating the $SU(d)$ sector of the 
matrix $\alpha$ with a corresponding sector for the quantum states, 
represented by transformations $R(\bar{S})$ in the adjoint representation 
$\bar{S} T^a \bar{S}^{-1} =\hat{n}_a \cdot\vec{T}$, 
$\hat{n}_a=R(\bar{S})\hat{e}_a$. 
In this manner, the matrices $e^{i 2\pi n/d}\,\bar{S}$ are mapped to the 
same point $R(\bar{S})$. In other words, a part of the evolution 
can be parametrized as $R(t)\in {\rm Adj}(d)$, or equivalently, in terms 
of a time dependent frame $\hat{n}_a(t)$. Note that for qubits the adjoint 
representation corresponds to $SO(3)$, the manifold used in ref. \cite{remy,milman} 
to describe maximally entangled states.
An evolution $\bar{S}(t)$ starting at $\bar{S}(0)$ and ending at 
$e^{i 2\pi/d}\bar{S}(0)$ defines an open path in $SU(d)$ and a topologically 
nontrivial closed path $R(t)\in {\rm Adj}(d)$. 
If this 
cyclic evolution were composed $d$ times, we would get a trivial path in 
${\rm Adj}(d)$, so that the number of nonequivalent classes is given by $d$.

The total phase can be written as
\begin{eqnarray}
\phi_{tot}&=&\arg{\{Tr[\mathbf{\alpha}^{\dagger}(0)\mathbf{\alpha}(t)]}\} = \phi_A + \phi_B 
\nonumber\\
&&
+\arg{\{Tr[\mathbf{\alpha}^{\dagger}(0)\bar{U}_A(t)\mathbf{\alpha}(0)\bar{U}^{\intercal}_B(t)]}\}\;,
\label{phitot}
\end{eqnarray}
while the dynamical phase is, 
\begin{eqnarray}
\phi_{dyn}=-i\,\int_0^t\,dt^{\prime}\,Tr[\mathbf{\alpha}^{\dagger}(t^{\prime})
\dot{\mathbf{\alpha}}(t^{\prime})]=\phi_A +\phi_B &&
\nonumber\\
-i\,\int_0^t\,dt^{\prime}\,Tr[\rho_B(0)\,\bar{U}_A^{\dagger}
\dot{\bar{U}}_A + \rho^{\intercal}_A(0)\, \dot{\bar{U}}_B^{\intercal}\bar{U}^\ast_B]\;,&&
\label{phidyn}
\end{eqnarray}
where $\rho_A=(S^\dagger Q^2 S)^\ast$, $\rho_B=Q^2$. 
For cyclic evolutions we have $\bar{U}_j(\tau)=e^{i2\pi n_j/d}\,\bar{U}_j(0)$.
Then, the total generated phase is $\phi_{tot} =\phi_A +\phi_B + 2\pi n/d$, $n=n_A+n_B$, where 
the values $n\neq 0, d, 2d, \dots$, correspond to topologically nontrivial paths. As already 
discussed, the total phase is always written as a dynamical plus a geometric part. In order to 
consider a fractional phase as topological, it must be built only as a part of the geometric 
phase, receiving no relevant contribution from the dynamical part. This means that at any 
time $t$, $0 \leq t \leq \tau$, we must have,
\begin{equation}
\int_0^t\,dt^{\prime}\,Tr[\rho_B(0)\,\bar{U}_A^{\dagger}
\dot{\bar{U}}_A + \rho^{\intercal}_A(0)\, \dot{\bar{U}}_B^{\intercal}\bar{U}^\ast_B]=0\;.
\label{dyn-term}
\end{equation}
This is satisfied by the maximally entangled states, for every possible local evolution $\bar{U}_j$. 
In this regard, the invariant quantities in the evolution can be written as $Tr[(Q^2)^p]$. 
In terms of the concurrence we can write 
\begin{equation}
Q^2= (1/d)\, I+ \sqrt{{\cal C}_m^2-{\cal C}^2}\,\,\hat{q}\cdot \vec{T}\;,
\label{Q2}
\end{equation}
where ${\cal C}_m=\sqrt{2(d-1)/d}$.
The ${\cal C}=0$ value corresponds to separable states. For maximally entangled 
states ${\cal C}={\cal C}_m$, giving $Q^2=(1/d)I$, and 
$\rho_A=\rho_B=(1/d)I$. In addition, for any $\bar{U}_j \in SU(d)$, the matrices 
$\bar{U}_j^{\dagger}\dot{\bar{U}}_j$ are combinations of the generators $T_a$. 
Therefore, using this information, the trace in the integrand of eq. (\ref{dyn-term}) 
vanishes.

Now, let us consider an evolution on the first qudit $A$. In this case,
$
\langle\psi(0)|\psi(t)\rangle =Tr[Q^{2}(0)\,\bar{U}_A(t)]
$, 
while the dynamical phase is, 
\begin{equation}
\phi_{dyn}=
\phi_A-i\,\int_0^t\,dt^{\prime}\,Tr[Q^2(0)\,\bar{U}_A^{\dagger}(t^{\prime})
\dot{\bar{U}}_A(t^{\prime})]\;.
\end{equation}
These phases do not depend on $\bar{S}(0)$ so that for simplicity we can consider 
$\bar{S}(0)=I$, that is, $\bar{U}_A(t)=\bar{S}(t)$.
For qubits $T_{a}=\sigma_{a}/2$ ($a=1,2,3$), where $\sigma_{a}$ are the Pauli 
matrices. We shall assume that the basis is chosen so that $\hat{q}(0)=\hat{e}_3$, that is,
$Q^2=I/2+\sqrt{1-{\cal C}^2}\,\sigma_{3}/2$. 
The unitary sector of the state evolution can be put in terms of Euler angles, in the form 
$\bar{U}_A(t)=U_m(t)\,V_3(t)$, where
\begin{equation}
U_m= e^{-i\varphi T_3} e^{i\theta T_2}  e^{i\varphi T_3}
\makebox[.3in]{,}
V_3=e^{i\chi T_3}.
\label{paramet}
\end{equation}
Note that for cyclic evolutions, $U_m(0)=U_m(\tau)$, while $V_3(0)= \pm V_3(\tau)$. 
In addition, $U_m$ can be expanded in terms of $I$, $T_1$ and $T_2$, as the term 
proportional to $T_3$ is obtained from 
$Tr[T_3\, e^{-i\varphi T_3} e^{i\theta T_2}  e^{i\varphi T_3}]= Tr[T_3\, e^{i\theta T_2}]=0$. 
Here, we have used that the latter exponential is a combination of $I$ and $T_2$.
Using a similar expansion for $e^{i\varphi T_3}$, we arrive to the conclusion that 
the terms in $U_m$ proportional to $T_1$, $T_2$ do not contribute to 
$\langle\psi(0)|\psi(t)\rangle$.

With the ingredients above we can work out the expression for the time 
evolving overlap 
\begin{equation}
\langle\psi(0)|\psi(t)\rangle = e^{i\phi_A}\cos\frac{\theta}{2}
\left[\cos\frac{\chi}{2}
+ i \sqrt{1-{\cal C}^2}\sin\frac{\chi}{2}\right]\;. 
\end{equation}
In terms of $Q^2$, $U_m$, and $V_3$, the dynamical phase is 
\begin{eqnarray}
\phi_{dyn}&=&\phi_A-i\,\int_0^t\,dt^{\prime}\,\frac{1}{2}
Tr[(I+\sqrt{1-{\cal C}^2}\,\sigma_3)
\nonumber\\
&\times&({U^\dagger_m} \dot{U}_m 
+ V_3^{\dagger}\dot{V}_3)]\label{phidyn2}\;.
\end{eqnarray}
Using $\dot{V}_3= i\,(\dot{\chi}/2)\,\sigma_3\,V_3$ and 
defining the unit vectors $\hat{m}_a$ so that
$
U_m\,\sigma_a\,U_m^{\dagger}=\hat{m}_a\cdot\vec{\sigma},
$
we get 
\begin{eqnarray}
\phi_g &=& \arctan\left[\sqrt{1-{\cal C}^2}\,\tan(\chi/2)\right] - 
\sqrt{1-{\cal C}^2}\,(\chi/2) 
\nonumber\\
&+& \sqrt{1-{\cal C}^2}\, (\Phi/2),
\label{mainresult}
\end{eqnarray}
with
$
\Phi \equiv \int_{0}^{t}\,dt^{\prime}\,
\hat{m}_1\cdot\dot{\hat{m}}_2\;.
\label{Phi}
$
In the last term, the frame $\hat{m}_a$ depends on $\theta \in [0, \pi)$ and 
$\varphi \in [0, 2\pi]$ defining a point on $S^2$, the surface 
of a sphere with unit radius. Then, $\hat{m}_a(\theta, \varphi)$ is a mapping 
$S^2 \rightarrow \hat{m}_a$, and the evolution on this sector is given by a curve, 
defined by $\theta(t),\varphi(t)$, contained on $S^2$. 
In this regard, for a cyclic evolution, 
one easily shows that $\Phi= \Omega$, where $\Omega$ is the solid 
angle subtended by the closed path \cite{cho3,lucho}. 
This term can be associated to the usual Berry phase for a single qubit. 

\begin{figure}
\begin{center} 
\includegraphics[scale=.35]{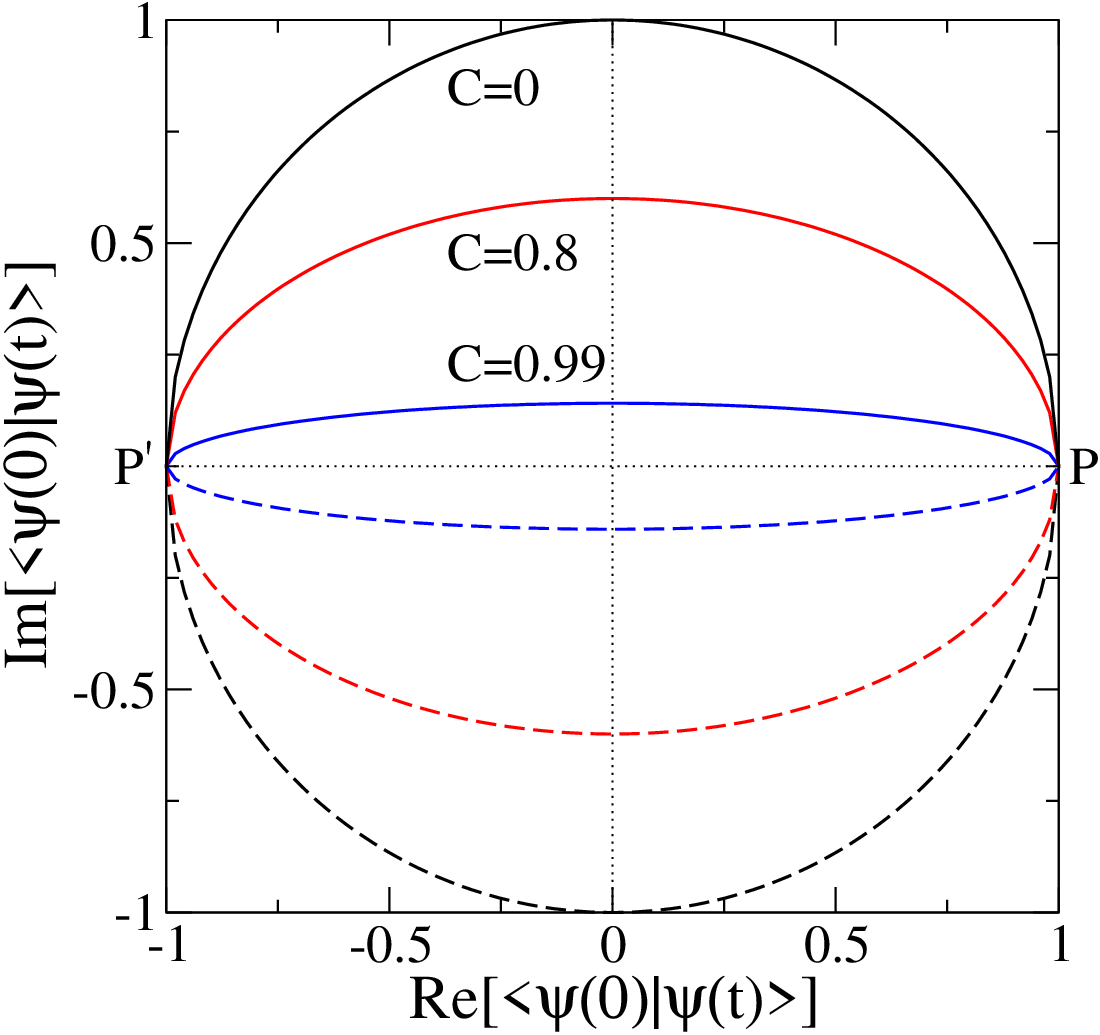}
\end{center} 
\caption{Time evolution of the quantum state overlap 
for a pair of qubits with different concurrences.}
\label{qubit}
\end{figure}

For a general evolution, we see that for product states (${\cal C}=0$), the first two terms in 
eq. (\ref{mainresult}) cancel each other 
while the last term coincides with the 
one given by the usual picture of the Bloch sphere evolution 
of a single qubit. On the other hand, for maximally entangled states (${\cal C}=1$), 
the last two terms vanish while the first term can assume 
only two discrete values $0$ or $\pi$. 
In fig.\ref{qubit} this evolution is represented as paths in the complex plane, 
where the overlap $\langle \psi(0)|\psi(t)\rangle$ is plotted 
for different values of the concurrence. 
This path degenerates to a circle 
for product states and to a straight line on the real axis as the 
concurrence approaches its maximum value $C=1$. It gives a graphical 
picture of the phase jump between $0$ and $\pi$ discussed in ref. \cite{milman}. 
This jump occurs when the evolving state crosses the subspace orthogonal to the initial one. 
Note that the solid lines in fig.\ref{qubit} correspond to closed paths 
since points $P$ and $P^{\prime}$ represent physically equivalent 
quantum states. Dashed lines correspond to additional closed paths.  

%
%

Now, let us study a simple nontrivial path that generalizes the $V_3$-sector for qubits 
(cf. eq. (\ref{paramet})) to the case of qudits. Consider an evolution of the form 
$\bar{U}_A(t)=V_N(t)$,
\begin{equation}
V_N(t)=e^{i\chi(t) E}
\makebox[.3in]{,}
\chi(0)=0\;,
\label{paramet-n}
\end{equation}
where $E$ is a diagonal traceless matrix with components,
\begin{equation}
E_{\alpha \alpha} =\left\{ \begin{array}{ll}
(1/d),  & \alpha =1, \dots, d-1 \\
(1/d)-1,& \alpha =d\;.
\end{array}\right. 
\label{diagE}
\end{equation}
This matrix can be written in terms of the $N$'th generator of $SU(d)$, $N=d^2-1$: 
$E={\cal C}_m T_N$.
When $\chi(\tau)=2\pi$, it is simple to see that $V_N(\tau)=e^{i2\pi/d}\,I$.
In the case where $\hat{q}(0)=\hat{e}_N$, we have,
\begin{equation}
Q^2(0)= (1/d)\, I+ \sqrt{1-({\cal C}/{\cal C}_m)^2}\, E\;.
\label{Qcuad}
\end{equation}
By expanding the exponential in eq. (\ref{paramet-n}) and 
using eq.(\ref{Qcuad}) we get,
\begin{equation}
\langle\psi(0)|\psi(t)\rangle = {\cal A}\, e^{i\chi/d} + {\cal B}\, e^{i(1-d)\chi/d}\;,
\end{equation}
with 
${\cal A}= \frac{d-1}{d}+\frac{1}{2}\sqrt{{\cal C}_m^2-{\cal C}^2}$ and 
${\cal B}= 1-{\cal A}$. 
Using ${V}_N^\dagger\, \dot{V}_N= i\,\dot{\chi}\,E$ in the dynamical phase, 
we arrive at 
\[
\phi_{g}=\arctan \left[\frac{{\cal A} \sin \frac{\chi}{d}+{\cal B} \sin \frac{(1-d)\chi}{d}}{{\cal A} 
\cos \frac{\chi}{d}+{\cal B} \cos \frac{(1-d)\chi}{d}}\right]
-\sqrt{{\cal C}_m^2-{\cal C}^2 }\,\frac{\chi}{2} \;.
\]
In the above example, for maximally entangled states, when $d\geq 3$ the total phase 
changes continuously from $0$ to $2\pi/d$, and the evolving state never becomes orthogonal 
to the initial state. This is in contrast to what happens in the $d=2$ case.
The minimum value for 
$|\langle\psi(0)|\psi(t)\rangle |^2$ is $({\cal A}-{\cal B})^2=(\frac{d-2}{d})^2$, 
attained when $\chi=\pi$. For $d=3$, the minimum overlap is $(1/3)^2$. 

It is interesting to look for topologically nontrivial evolutions for qudits with similar 
properties to those displayed by qubits. In the $d=3$ case, this can be realized as follows. 
Let us consider the path $\bar{U}_A(\chi(t))$, continuously evolving from $\bar{U}_A(0)=I$ to
$\bar{U}_A(2\pi)=e^{i2\pi/3} I$, defined by a diagonal unitary matrix with nontrivial 
elements $e^{i\phi_\alpha}$ such that $\phi_1=2\chi/3+[2(\pi-\zeta)/3]\Theta(\chi-\pi)$, 
$\phi_2=-2\chi/3$, and $\phi_3=-(\phi_1+\phi_2)$; $\Theta(\chi)$ is the Heaviside 
function.
For maximally entangled states, we have
\[
\langle\psi(0)|\psi(t)\rangle =\left\{ \begin{array}{ll}
\frac{1}{3} [1+ 2\cos (\frac{2\chi}{3})]\;,  & \chi \in [0,\pi] \\
\frac{1}{3} [1+ 2\cos (\frac{2(\chi+\pi)}{3})]\, e^{i\frac{2\pi}{3}}\;,& \chi \in [\pi,2\pi]\;.
\end{array}\right. 
\]
Then, we see that the total phase vanishes in the first part of the evolution, 
while it takes the fractional value $2\pi/3$ in the second part. In addition, 
at $\chi =\pi$, when the phase changes discontinuously, the state $|\psi(t)\rangle$ 
becomes orthogonal to the initial state.

\begin{figure}[h]
\includegraphics[scale=.4]{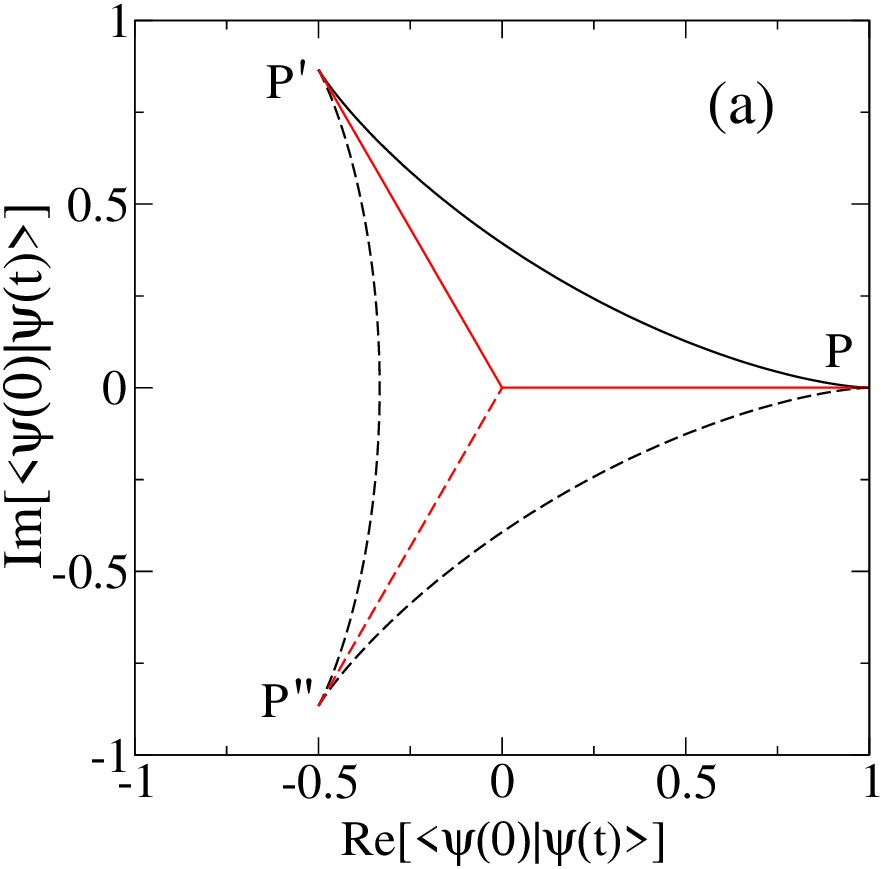}
\includegraphics[scale=.4]{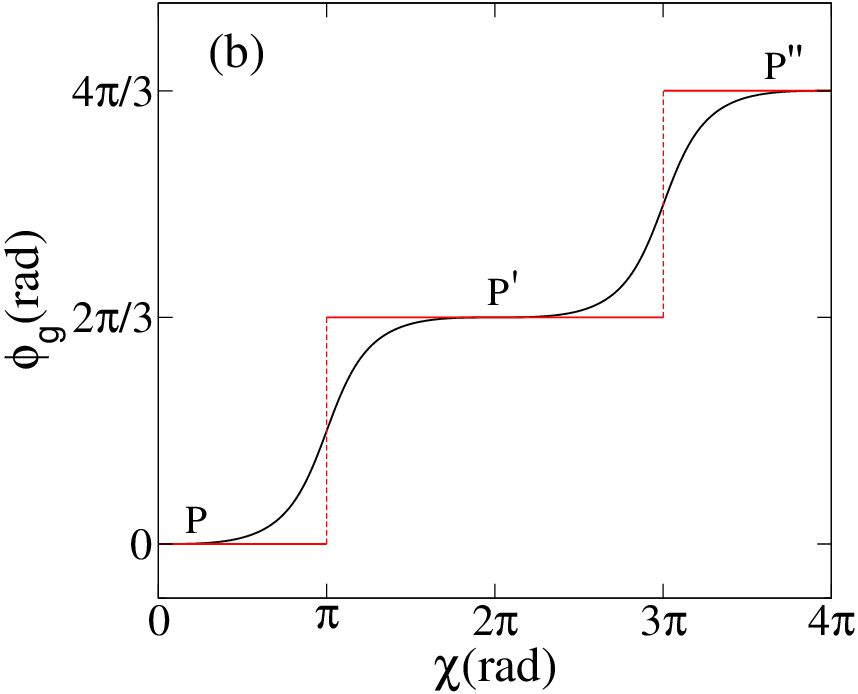}
\caption{(a) Complex plane representation of the quantum state overlap 
for a pair of qutrits with maximal concurrence. Two different time evolutions 
are considered. (b) The corresponding stepwise evolution of the geometric phase.}
\label{qutrit}
\end{figure}
%

Both qutrit evolutions are represented in fig.\ref{qutrit}a, 
where the overlap $\langle \psi(0)|\psi(t)\rangle$ is plotted in 
the complex plane for maximal concurrence. The first cyclic evolution 
from $P$ to $P^{\prime}$ is represented by the solid black line 
clearly showing that the overlap between 
the initial and the evolving quantum states never vanishes. On the 
other hand, the second evolution (red online) shows a path 
crossing the origin of the complex plane, where the evolving quantum 
state becomes orthogonal to the initial one. 
The dashed lines correspond to additional closed 
paths defining three vertices which evidence the fractional 
phase values. In fig.\ref{qutrit}b, we plot the associated geometric 
phase evolution, showing a stepwise behavior with two jumps between 
the fractional values $0$, $2\pi/3$, and $4\pi/3$. For the first 
evolution (black) smooth jumps occur, while for the second evolution 
(red online) they are discontinuous. 

As a conclusion, in this letter we studied unitary local operations 
on a pair of qudits, showing that fractional phases naturally appear 
when cyclic evolutions are considered. These fractional values are 
related to different homotopy classes of closed paths in the two-qudit 
Hilbert space. 
The geometric phase has been calculated in terms of the I-concurrence 
introduced in ref.\cite{Iconc}. In the case of maximally entangled states, 
the fractional values originate solely from the geometric part of the 
phase evolution, since the dynamical part vanishes at all times. 

The fractional phase of maximally entangled states is 
built in a stepwise evolution, where the phase jumps 
between discrete values in steps of $2\pi/d$. 
For qubits this jump is strictly discontinuous, while for 
qutrits, it may be discontinuous or not, depending on the 
particular evolution considered. 
Due to its stepwise evolution, we expect 
the fractional phase acquired by maximally entangled qudits to be 
particularly robust against the influence of the environment. In order to 
produce a relevant change, any external 
noise would have to cause a large fluctuation, driving 
the two-qudit system through a phase step. Since the phase jump for 
qubits is strictly discontinuous, its robustness should be even more 
pronounced. These results can be important to proposals of quantum 
gates based on topological phases.

\section*{Acknowledgements}
We are grateful to M. Sarandy, P. Milman and E. Sj\"oqvist for useful 
discussions. 
Funding was provided by Coordena\c c\~{a}o de Aperfei\c coamento de 
Pessoal de N\'\i vel Superior (CAPES), Funda\c c\~{a}o de Amparo \`{a} 
Pesquisa do Estado do Rio de Janeiro (FAPERJ-BR), and Instituto Nacional 
de Ci\^encia e Tecnologia de Informa\c c\~ao Qu\^antica (INCT-CNPq).

\end{document}